\documentclass[10pt]{iopart}

\usepackage{cite}

\def\be{\begin{equation}}
\def\ee{\end{equation}}
\def\bea{\begin{eqnarray}}
\def\eea{\end{eqnarray}}
\def\lb{\label}

\def\Om{\Omega}
\def\l{\lambda}              
\def\bx{\mbox{\boldmath $x$}}
\def\ba{\mbox{\boldmath $a$}}
\def\bb{\mbox{\boldmath $b$}}

\def\bn{\mbox{\boldmath $n$}}
\def\x0{x_{(0)}(\lambda )}

\begin{document}

\title[World function and time transfer]{World function and time transfer: general post-Minkowskian expansions}

\author{Christophe Le Poncin-Lafitte\dag\footnote[3]{To
whom correspondence should be addressed (leponcin@danof.obspm.fr)}, Bernard Linet\ddag\ and Pierre Teyssandier \dag}

\address{\dag\ D\'{e}partement Syst\`{e}mes de R\'{e}f\'{e}rence Temps et Espace, CNRS/UMR 8630\\
Observatoire de Paris, 61 avenue de l'Observatoire, F-75014 Paris, France}
\address{\ddag\ Laboratoire de Math\'ematiques et Physique Th\'eorique, CNRS/UMR 6083 \\
Universit\'e Fran\c{c}ois Rabelais, F-37200 Tours, France \\}

\begin{abstract}

In suitably chosen domains of space-time, the world function may be a powerful tool for modelling the deflection of light and the time/frequency transfer. In this paper we work out a recursive procedure for expanding 
the world function into a perturbative series of ascending powers of the Newtonian gravitational 
constant $G$. We show rigorously that each perturbation term is given by a line integral taken
along the unperturbed geodesic between two points. Once the world function is known, it becomes possible 
to determine the time transfer functions giving the propagation time of a photon between its emission and its reception. We establish that the direction of a light ray as measured in the 3-space relative to an 
observer can be derived from these time transfer functions, even if 
the metric is not stationary. We show how to derive these functions 
up to any given order in $G$ from the perturbative expansion of the world 
function. To illustrate the method, we carry out the calculation of the world function and 
of the time transfer function outside a static, spherically symmetric body up to the order $G^2$, 
the metric containing three arbitrary parameters $\beta$, $\gamma$, $\delta$.

\end{abstract}

\maketitle

\section{Introduction}

With advances in technology, it will become indispensable to determine some of the relativistic effects 
in the propagation of light beyond the first order in the Newtonian gravitational constant $G$, in particular 
in the area of space astrometry. The Global Astrometric Interferometer for Astrophysics (GAIA) \cite{perryman} and the Space Interferometric Mission (SIM) \cite{danner} are already planned to measure the positions and/or the parallaxes of celestial objects with typical uncertainties in the range 1-10 $\mu$arcsecond ($\mu$as). The Laser Astrometric Test Of Relativity (LATOR) mission is an even more ambitious program since its 
primary objective is to measure the bending of light near the Sun to an accuracy of 0.02 $\mu$as \cite{turyshev}. With such requirements in experimental precision, it is clear that the effects of the second order in $G$ must be taken into account in the near future since they result in several $\mu$as deflection for a light ray grazing the Sun \cite{epstein, fischbach, richter1, nordtvedt}.

In order to obtain a convenient modelling of the above-mentioned projects, it is necesssary to determine the
deflection of light rays between two points $x_A$ and $x_B$ of space-time.  
In almost all of the theoretical studies devoted to this problem, the properties of light rays are determined by integrating the differential equations of the null geodesics. This procedure is workable as long as one contents oneself with analyzing the effects of first order in $G$, as it is proven by the generality of the results 
obtained in \cite{klioner1, klioner2, kopeikin1, kopeikin2, kopeikin3, ciufolini1, ciufolini3, klioner3}. Unfortunately, analytical solution of the geodesic equations requires cumbersome calculations when terms of second order in $G$ are taken into account, even in the case of a static, spherically symmetric 
space-time \cite{epstein, fischbach, richter1, richter2}. The goal of the present paper is to develop a procedure allowing to determine the deflection of light and the travel time of photons without integrating the geodesic 
equations. 

We assume that space-time is globally regular with the topology $I\! \! R \times I \! \! R^{\,3}$, i.e. without event horizon. According to these assumptions, we admit that the Lorentzian metric is represented at any point $x$ by a series in ascending powers of $G$
\be \lb{1n}
g_{\mu\nu}(x, G) = g^{(0)}_{\mu\nu} + \sum_{n=1}^{\infty}G^n \, g_{\mu\nu}^{(n)}(x) \, ,
\ee 
\noindent
where 
\[
g^{(0)}_{\mu\nu} = \eta_{\mu\nu} = \mbox{diag} \, (1, -1, -1, -1) \, 
\]
\noindent
in any quasi Cartesian coordinates. Neglecting all terms involving $g_{\mu\nu}^{(n+1)}$, $g_{\mu\nu}^{(n+2)}$, ... defines the so-called $n$th post-Minkowskian approximation. 

Essentially, our method uses the notion of world function as developed by Ruse \cite{ruse1} and Synge \cite{synge1, synge2}. The world function $\Omega (x_A,x_B)$ is defined as half the squared geodesic distance between two points $x_A$ and $x_B$. Before entering into the presentation of the method, it must be noted that this definition raises a serious problem, since there could exist points $x_A$ and $x_B$ connected by several geodesics. Given a point $x_A$, it is usual to define the normal convex neighbourhood of $x_A$, say ${\cal N}_{x_{A}}$, as the set of all points $x$ joined to $x_A$ by a unique geodesic path \cite{poisson}. The world function $\Omega(x_A, x)$ is then well defined for 
$x\in {\cal N}_{x_{A}}$. In agreement with the assumptions mentioned in the previous paragraph, we suppose that if $x_B$ is joined to $x_A$ by a null geodesic, then this geodesic is unique. Denoting the light cone at $x_A$ by ${\cal C}_{x_{A}}$, this means that ${\cal C}_{x_{A}}\subset {\cal N}_{x_{A}}$. Thus we can use the world function to study the deflection of the light and the travel time of photons between $x_A$ and $x_B$. Let us point out that our assumptions are compatible with the existence of multiple-images effects involving different travel times for the photons (see, e.g., \cite{giannoni1} and references therein). 

We shall see in section 2 that the invariant two-point function $\Omega (x_A, x_B)$ has very interesting properties. First 
of all, considering the geodesic path connecting $x_A$ and $x_B$, it may be shown that the tangent vectors at $x_A$ and $x_B$ are provided by the gradients of the world function. Second, it 
is clear that $x_A$ and $x_B$ are joined by a null geodesic if and only if $\Omega (x_A, x_B) = 0$. 
As a consequence, the deflection of a light ray emitted at a finite distance may be easily determined 
when the world function of a given space-time is known, just as the travel time of a photon between 
two points. This means that the world 
function allows to calculate the time/frequency transfer functions and can be extremely useful in situations involving a comparison between two distant clocks, as it has been shown by two of us in a recent work \cite{linet1} extending the results obtained in \cite{ashby} and \cite{blanchet1}. Other 
possible applications deserve to be emphasized. For example, the quantum phase accumulated by a 
freely falling particle is proportional to the geodesic distance covered through 
space-time \cite{stodolsky}. So, the world function can be relevant for modelling 
situations like neutrino oscillations \cite{linet2}. 

Taking equation (\ref{1n}) into account, it is possible to assume that $\Omega (x_A, x_B)$ admits an expansion as follows
\be \lb{4nn}
\Omega(x_A, x_B) = \Omega^{(0)}(x_A, x_B) + \sum_{n=1}^{\infty} G^n \Omega^{(n)}(x_A, x_B)  \, .
\ee
\noindent
The primary objective of the paper is to yield a recursive procedure for determining each term 
$\Omega^{(n)}(x_A, x_B)$ within the $n$th post-Minkowskian approximation without calculating the geodesic joining $x_A$ and $x_B$.

Our method is inspired by a procedure found by Buchdahl \cite{buchdahl4} for obtaining the perturbative expansion of the characteristic function defined as $V= \sqrt{2\mid \!\Omega \!\mid}$. However, the gradients of $V(x_A, x_B)$ are not defined when $x_A$ and $x_B$ are linked by a null geodesic. So we prefer developing a new procedure involving $\Omega$ itself. Considering that $x_A$ and $x_B$ are fixed points, we use the Hamilton-Jacobi equation satisfied by $\Omega (x_A, x)$ at point $x$. It is easily seen that this equation may be replaced by an infinite set of ordinary differential equations 
for the perturbation terms $\Omega^{(n)}(x_A, x)$ when $x$ is constrained to move along the zeroth-order geodesic connecting $x_A$ and $x_B$. As a consequence, each term $\Omega^{(n)}(x_A, x_B)$ is obtained in the form 
of a line integral along a straight line in the background Minkowski metric. Thus, we obtain a recursive 
procedure which completely avoids the calculation of the perturbation of the geodesic joining the given end points. 

Within the first post-Minkowskian approximation, we recover a simple integral expression of $\Omega^{(1)}$ already given by Synge \cite{synge3}. In contrast, the fact that one has to calculate line integrals with integrands containing products of the first-order derivatives of terms $\Omega^{(n-p)}$, $p = 1$, ..., $n-1$ proves to be awkward in attempts at getting explicit expressions of $\Omega^{(n)}$ for $n\geq 2$. Fortunately, at least in the case where $n=2$, we obtain a simplified formula for $\Omega^{(2)}$ where the quadratic terms in the first derivatives of $\Omega^{(1)}$ appear only as boundary terms. Then, we develop an iterative procedure allowing to deduce the propagation time of a photon correct to within $(n+1)$th-order terms from the expressions of $\Omega^{(n)}$, $\Omega^{(n-1)}$,..., $\Omega^{(0)}$. 
 
Finally, we derive the world function and the propagation time of light in a static, spherically symmetric 
space-time within the second post-Minkowskian approximation. In order to get expressions valid for a wide class 
of gravitational theories, we assume that the metric involves three arbitrary parameters: the usual post-Newtonian parameters $\beta$ and $\gamma$ \cite{will1}, and a supplementary parameter $\delta$, introduced to take into account the term of order $G^2$ in spatial components $g_{ij}$ (see, e.g., \cite{damour} and references therein). Complete results were obtained by John \cite{john1} involving quite cumbersome calculations. Without any reference to this earlier paper, a complicated integration of the 
geodesic equations was performed by Richter and Matzner \cite{richter2} in order to calculate the time delay for a special coordinate system adapted to the direction of the light ray at the emission point. More recently, Brumberg \cite{brumberg} has used the Euler-Lagrange equations of null geodesics to find the propagation time of 
light as a function of the boundary conditions in the second post-Minkowskian approximation of the Schwarzschild metric. For our part, we do not need determining the null geodesics to get the world function and the propagation time of light and we use only elementary quadratures to perform the calculations. 
  
The paper is organized as follows. In section 2 the general definition of 
the world function and its fundamental properties are recalled. In
section 3 we introduce two distinct time transfer functions and we give their 
general relations with the world function. We establish that all the calculations 
relative to the direction of a light ray or to the frequency transfers can be 
performed when one of the two time transfer functions is explicitly known. In section 4 we set out 
the recursive method which allows to determine the world function in the $n$th post-Minkowskian 
approximation for any $n \geq 2$. In section 5 we show how to deduce the time transfer functions from the world 
function at any post-Minkowskian order of approximation. In 
section 6 we determine explicitly the world function and the time transfer outside a 
static, spherically symmetric body within the second post-Minkowskian approximation. Some concluding remarks are given in section 7. In Appendix A we give an alternative derivation of the integral expression for 
$\Omega^{(2)}(x_A,x_B)$, based on an integration of the Euler-Lagrange equations of geodesics.  
  
In this paper $c$ is the speed 
of light in a vacuum. The Lorentzian metric of space-time $V_4$ is denoted by $g$. 
The signature adopted for $g$ is $(+---)$. We suppose that space-time is 
covered by some global quasi-Galilean coordinate system $(x^\mu)=(x^0,\bx )$, where $x^0=ct$,
$t$ being a time coordinate, and $\bx=(x^i)$. We assume that the curves of equations $x^i$ = const 
are timelike, which means that $g_{00}>0$ anywhere. We employ the vector notation $\ba$ in 
order to denote $(a^1,a^2,a^3)=(a^i)$. Considering two such quantities
$\ba$ and $\bb$ we use $\ba \cdot \bb$ to denote $a^ib^i$ (Einstein convention on 
repeated indices is used). The quantity $\vert \ba \vert$ stands for the ordinary Euclidean 
norm of $\ba$. For any quantity $f(x^{\lambda})$, $f_{, \alpha}$ denotes the partial derivative 
of $f$ with respect to $x^{\alpha}$. The indices in parentheses characterize the order of 
perturbation. They are set up or down, depending on the convenience.

\section{Definition and fundamental properties of Synge's world function}

Since Synge's world function is not widely employed, we 
briefly recall how its fundamental properties can be straightforwardly derived 
from the variational principle defining the geodesic curves (for the characteristic 
function $ V = \sqrt{2\mid \!\Omega \!\mid}$, see e.g. Buchdahl 
\cite{buchdahl1, buchdahl2, buchdahl3}). We begin with 
recalling some useful results concerning the general variation of a 
functional defined as an action in classical mechanics. 

Let $x_A$ and $x_B$ be two points of space-time connected by a 
differentiable curve $C_{AB}$ defined by the parametric equations 
$x^{\alpha}=x^{\alpha}(\zeta)$, with $\zeta_A \leq \zeta \leq \zeta_B$. 
Given a Lagrangian function $L(x^{\alpha}, \dot{x}^{\beta})$ with 
$\dot{x}^{\beta}(\zeta)=dx^{\beta}(\zeta)/d\zeta$, one can define the 
functional $\widetilde{S}$ by 
\be \lb{01}
\widetilde{S}[C_{AB}]=\int_{\zeta_A}^{\zeta_B}L(x^{\alpha}(\zeta),
\dot{x}^{\beta}(\zeta))d\zeta \, , 
\ee
\noindent
the integral being taken along $C_{AB}$. Let us consider now an arbitrarily neighbouring curve 
$C'_{A'B'}$ connecting points $x_{A'}$ and $x_{B'}$, represented by parametric 
equations $x'^{\alpha}=x'^{\alpha}(\zeta ')$, with $\zeta_{A'} \leq \zeta ' \leq \zeta_{B'}$. 
Let $x^{\alpha}_A+\delta x^{\alpha}_{A}=x^{\alpha}_{A'} $ and 
$x^{\alpha}_{B}+\delta x^{\alpha}_{B}=x^{\alpha}_{B'}$ be the 
coordinates of $x_{A'}$ and $x_{B'}$ respectively. Putting 
$\bar{\delta}x^{\alpha}(\zeta)=x'^{\alpha}(\zeta)-x^{\alpha}(\zeta)$, one can 
define a one-to-one correspondance between $C_{AB}$ and $C_{A'B'}$ by the 
infinitesimal transformation 
\be \lb{02}
x^{\alpha}(\zeta )\rightarrow x'^{\alpha}(\zeta +\delta \zeta)=
x^{\alpha}(\zeta )+\delta x^{\alpha}(\zeta ) \, , 
\ee
\noindent
where
\be \lb{03}
\delta x^{\alpha}(\zeta)=\bar{\delta}x^{\alpha}(\zeta )+
\dot{x}^{\alpha}(\zeta ) \delta \zeta  
\ee
\noindent
with the boundary conditions
\be \lb{04}
\delta x^{\alpha}(\zeta_A)=\delta x^{\alpha}_{A} \, , \quad 
\delta x^{\alpha}(\zeta_B)=\delta x^{\alpha}_{B} \, . 
\ee

Performing an integration by parts leads directly to the following expression 
for the quantity $\delta \widetilde{S}=\widetilde{S}[C_{A'B'}]-\widetilde{S}[C_{AB}]$
(see, e.g., \cite{gelfand}) 
\be \lb{05}
\delta \widetilde{S}=\int_{\zeta_A}^{\zeta_B}\left[ \frac{\partial L}
{\partial x^{\alpha}}-
\frac{d}{d\zeta} \left( \frac{\partial L}{\partial \dot{x}^{\alpha}}\right) 
\right] \bar{\delta}x^{\alpha} (\zeta )d\zeta + 
\left[ p_{\alpha}\delta x^{\alpha}-H\delta \zeta \right]_{A}^{B} \, ,
\ee
\noindent
where $p_{\alpha}$ is the 4-momentum belonging to $x^{\alpha}$ and $H$ is the 
Hamiltonian: 
\be \lb{06}
p_{\alpha}=\frac{\partial L}{\partial \dot{x}^{\alpha}} \, , \quad 
H=p_{\alpha}\dot{x}^{\alpha}-L \, . 
\ee

In the following we assume that there exist domains ${\cal D}$ of space-time such 
that whatever $x_A \in {\cal D}$ and $x_B \in {\cal D}$, $x_A$ and $x_B$ are linked by a unique curve $\Gamma_{AB}$ realizing an extremum 
of the functional $\widetilde{S}$. Then, describing $\Gamma_{AB}$ by parametric equations 
$x^{\alpha}=x^{\alpha}(\zeta )$, we can associate to any domain $\cal D$ a function $S$ of
$x_A,x_B,\zeta_A,\zeta_B$ defined as 
\be \lb{07}
S(x_A,x_B,\zeta_A,\zeta_B)=\int_{\Gamma_{AB}}L\left( x^{\alpha} (\zeta ),
\dot{x}^{\beta}(\zeta ) \right) d \zeta \, . 
\ee
Extending a terminology currently used in mechanics, we shall call $S$ the characteristic (or principal) function belonging to the Lagrangian $L$ adapted to the domain $\cal D$. Since the functions 
$x^{\alpha}(\zeta )$ satisfy the Euler-Lagrange equations
\be \lb{08}
\frac{\partial L}{\partial x^{\alpha}}-\frac{d}{d \zeta} 
\left( \frac{\partial L}{\partial \dot{x}^{\alpha}}\right) = 0 \, , 
\ee
\noindent
the total variation of $S(x_A,x_B,\zeta_A,\zeta_B)$ reduces to the boundary 
terms in equation (\ref{05}). Consequently, one has the relations
\be \lb{09}
\frac{\partial S}{\partial x^{\alpha}_{A}} = - (p_{\alpha})_{A} \, , \quad
\frac{\partial S}{\partial x^{\alpha}_{B}} = (p_{\alpha})_{B} \,  ,
\ee 
\be \lb{010}
\frac{\partial S}{\partial \zeta_A} = H_A \, , \quad
\frac{\partial S}{\partial \zeta_B} = - H_B \, . 
\ee
Considering that $H$ is a function of $x^{\alpha}$ and $p_{\alpha}$, then 
substituting for $(p_{\alpha})_{A}$ and $(p_{\alpha})_{B}$ from equations (\ref{09}) 
into equations (\ref{010}), it is easily seen that $S(x_A,x_B,\zeta_A,\zeta_B)$ 
satisfies a Hamilton-Jacobi equation both at $x^{\alpha}_{A}$ and at 
$x^{\alpha}_{B}$
\be \lb{011}
\frac{\partial S}{\partial \zeta_A} =  H \left( x^{\alpha}_{A} , 
- \frac{\partial S}{\partial x^{\alpha}_{A}} \right) \, , \quad
\frac{\partial S}{\partial \zeta_B} = - H \left( x^{\alpha}_{B} , 
\frac{\partial S}{\partial x^{\alpha}_{B}} \right) \, . 
\ee 

Let us apply these results to space-time $V_4$ endowed with a Lorentzian metric $g_{\mu \nu}$ and insert 
in equation (\ref{01}) the Lagrangian defined by
\be \lb{021}
L=\frac{1}{2} g_{\alpha \beta}\dot{x}^{\alpha}\dot{x}^{\beta} \, .
\ee
\noindent
Each curve $\Gamma_{AB}$ parametrized by $\zeta$ for which $\widetilde{S}$ has an extremum is a geodesic path joining $x_A$ and $x_B$. Parameter $\zeta$ is said to be affine. According to (\ref{06}), we have 
now 
\be \lb{021aa}
p_{\alpha}= g_{\alpha \beta}\dot{x}^{\beta} \, , \quad H=\frac{1}{2} g^{\alpha \beta} p_{\alpha}p_{\beta} \, ,
\ee 
\noindent
from which it is easily deduced that $H=L$ on any geodesic path. Owing to the 
fact that $L$ does not contain $\zeta$ as an explicit variable, this last 
equality implies that the Lagrangian $L$ defined by (\ref{021}) is a constant 
of the motion.

We are only concerned here with a weak-field metric 
represented by equation (\ref{1n}) throughout space-time. So we henceforth restrict our attention to the domains 
${\cal D} = {\cal N}_{x_{A}}$, ${\cal N}_{x_{A}}$ being defined in Introduction. Then whatever 
$x_B \in {\cal N}_{x_{A}}$, there exists a unique geodesic path $\Gamma_{AB}$ connecting $x_A$ and $x_B$. Let us 
denote by $\widehat{\Omega}$ the characteristic function belonging to the Lagrangian (\ref{021}) adapted to 
${\cal N}_{x_{A}}$. According to (\ref{07}) we have
\be \lb{026l}  
\widehat{\Omega} (x_A,x_B,\zeta_A,\zeta_B)=\frac{1}{2} 
\int_{\zeta_A}^{\zeta_B} g_{\alpha \beta} \dot{x}^{\alpha} 
\dot{x}^{\beta}d\zeta \, , 
\ee
\noindent 
the integral being taken along $\Gamma_{AB}$. Since $L=H$ is a constant of 
the motion, we have $L=H_A=H_B$ on $\Gamma_{AB}$. Inserting these relations 
into equation (\ref{026l}) yields 
\be \lb{026aa}  
H_A = H_B = \frac{\widehat{\Omega}(x_A,x_B,\zeta_A,\zeta_B)}{\zeta_B-\zeta_A } \, .
\ee
\noindent 
Substituting these expressions of $H_A$ and $H_B$ into equations (\ref{010}), we get
\be \lb{028}
\frac{\partial \widehat{\Omega}}{\partial \zeta_A}= 
\frac{\widehat{\Omega}}{\zeta_B-\zeta_A} \, , \quad 
\frac{\partial \widehat{\Omega}}{\partial \zeta_B}=  
-\frac{\widehat{\Omega}}{\zeta_B-\zeta_A} \, . 
\ee
\noindent
Integrating these equations, we find that $\widehat{\Omega}$ may be written as
\be \lb{029}
\widehat{\Omega}(x_A,x_B,\zeta_A,\zeta_B)= 
\frac{\Omega (x_A,x_B)}{\zeta_B-\zeta_A} \, ,
\ee
\noindent
where $\Omega (x_A,x_B)$ is a function of $x_A$ and of $x_B$. This two-point function, 
which is symmetric in $x_A$ and $x_B$, is called the world function. Denoting by 
$\lambda$ the unique affine parameter such that $\lambda_A=0$ and $\lambda_B=1$, we infer from equations (\ref{026l}) and (\ref{029}) that $\Omega (x_A,x_B)$ may be written as 
\be \lb{d}
\Omega (x_A,x_B)=\frac{1}{2}\int_{0}^{1}g_{\mu \nu}(x^{\alpha}(\lambda ))
\frac{dx^{\mu}}{d\lambda}\frac{dx^{\nu}}{d\lambda}d\lambda \, ,
\ee
\noindent
the integral being taken along $\Gamma_{AB}$. It can be easily seen that
\be \lb{d1}
\Omega (x_A,x_B)=\frac{1}{2}\epsilon_{AB}[s_{AB}]^2 \, ,
\ee
\noindent
where $\epsilon_{AB} = 1, 0, -1$ when $\Gamma_{AB}$ is a timelike, a null or a spacelike geodesic, 
respectively, and $s_{AB}$ is the geodesic distance between $x_A$ and $x_B$, i.e.
$$
s_{AB} = \int_{\Gamma_{AB}} \sqrt{\vert g_{\mu\nu}dx^{\mu} dx^{\nu} \vert}  \, .
$$

The relevance of the world function in the problems related to the light deflection, the time delay 
or the gravitational frequency shift is justified by the 
following properties, which are easily deduced from the above-mentioned theory.  

\medskip
{\bf Property 1}.  {\em The covariant components of the vectors tangent 
to the geodesic path} $\Gamma_{AB}$ 
{\em at} $x_A$ {\em and} $x_B$ {\em respectively, are given by} 
\be \lb{p1a}
\left( g_{\mu \nu}\frac{dx^{\nu}}{d\lambda}\right)_A = -
\frac{\partial \Omega}{\partial x^{\mu}_{A}} (x_A, x_B)\, ,
\ee
\be \lb{p1b}
\left( g_{\mu \nu}\frac{dx^{\nu}}{d\lambda}\right)_B =
\frac{\partial \Omega}{\partial x^{\mu}_{B}}(x_A, x_B) \, .
\ee

These fundamental formulae are immediately deduced from equations (\ref{09}) and 
(\ref{021aa}). They show that the vectors tangent to the geodesic path 
$\Gamma_{AB}$ at $x_A$ and $x_B$ can be explicitly  determined when the world 
function $\Omega (x_A,x_B)$ is known.

Any other affine parameter $\zeta$ along $\Gamma_{AB}$ is such that 
\be \lb{q1}
\zeta = (\zeta_B - \zeta_A)\lambda + \zeta_A \, ,
\ee
\noindent
where $\zeta_A$ and $\zeta_B$ are the values corresponding to $x_A$ and $x_B$, respectively. 
As a consequence, we have for the tangent vector $dx^{\mu}/d\zeta$ the general formulae
\be \lb{q2a}
\left( g_{\mu \nu}\frac{dx^{\nu}}{d\zeta}\right)_A = - \frac{1}{\zeta_B - \zeta_A}\,
\frac{\partial \Omega}{\partial x^{\mu}_{A}}\,(x_A, x_B) \, ,
\ee
\be \lb{q2b} 
\left( g_{\mu \nu}\frac{dx^{\nu}}{d\zeta}\right)_B = \frac{1}{\zeta_B - \zeta_A}\,
\frac{\partial \Omega}{\partial x^{\mu}_{B}}\,(x_A, x_B) \, .
\ee

It is clear that point $x_B$ may be replaced in (\ref{q2b}) by any point $x(\zeta)$ on 
$\Gamma_{AB}$ which differs from $x_A$. Returning to $\lambda$ for the sake of simplicity, we find 
that the covariant components of the vector tangent to $\Gamma_{AB}$ at point $x(\lambda)$ are 
given by
\be \lb{q3}
\left( g_{\mu \nu}\frac{dx^{\nu}}{d\lambda}\right)_{x(\lambda)} = \frac{1}{\lambda}
\frac{\partial \Omega}{\partial x^{\mu}}(x_A, x(\lambda)) \, ,
\ee
\noindent
where $\partial \Omega/\partial x^{\mu}$ denotes the partial derivative of $\Omega(x_A, x)$ with respect to $x^{\mu}$ at point $x$.
It immediately follows from equations (\ref{q3}) that the system of equations    
\be \lb{q3a}
\left( \frac{dx^{\mu}}{d\lambda}\right)_{x(\lambda)} = \frac{1}{\lambda}\, g^{\mu \rho}(x(\lambda))
\frac{\partial \Omega}{\partial x^{\rho}}(x_A, x(\lambda)) \, ,
\ee
\noindent
can be regarded as the first-order differential system governing the geodesic paths passing through a given point $x_A$. The regularity of this system at $x_A$ is a direct consequence of the following property \cite{synge2}.

\medskip
{\bf Property 2}.  {\em The first-order partial derivatives of} $\Omega (x_A, x)$ 
{\em may be expanded as}
\be \lb{q4} 
\fl\frac{\partial \Omega}{\partial x^{\mu}}(x_A, x) = g_{\mu\nu}(x_A)(x^{\nu} - x^{\nu}_{A}) + 
C_{\mu \alpha \beta}(x_A, x)(x^{\alpha} - x^{\alpha}_{A})(x^{\beta} - x^{\beta}_{A}) \, ,
\ee
\noindent
{\em where the functions} $C_{\mu \alpha \beta}(x_A, x)$ {\em remain bounded in the neighbourhood of} $x_A$.

It results from this property that $\partial \Omega /\partial x^{\mu}(x_A, x) \rightarrow 0$ as 
$x \rightarrow  x_A$ and that the r.h.s. of equations (\ref{q3}) and (\ref{q3a}) remains bounded as 
$\lambda \rightarrow 0$. 

Now, the following statement can be straightforwardly derived from equations (\ref{011}), 
(\ref{021aa}) and (\ref{026aa}).

\medskip
{\bf Property 3}.  {\em The world function} $\Omega (x_A,x_B)$ {\em satisfies 
the Hamilton-Jacobi equations}
\be \lb{p2a}
\frac{1}{2}\, g^{\alpha\beta}(x_A)\frac{\partial \Omega}{\partial x^{\alpha}_{A}}(x_A, x_B) 
\frac{\partial \Omega}{\partial x^{\beta}_{A}}(x_A, x_B) =  \Omega (x_A,x_B)\, ,
\ee
\be \lb{p2b}  
\frac{1}{2}\, g^{\alpha \beta}(x_B)\frac{\partial \Omega}{\partial x^{\alpha}_{B}}(x_A, x_B) 
\frac{\partial \Omega}{\partial x^{\beta}_{B}}(x_A, x_B) =  \Omega (x_A,x_B)\, . 
\ee

As we shall see below, these equations and Property 2 enable to construct the world function in any post-Minkowskian approximation. 

Properties 1, 2 and 3 are valid whatever the nature of the geodesic curve joining 
$x_A$ ant $x_B$. In the case of null geodesics, (\ref{d}) and $L=H_A=H_B = 0$ 
immediately lead to the following statement.

\medskip
{\bf Property 4}.  {\em Two points} $x_A$ {\em and} $x_B$ {\em are joined 
by a light ray if and only if the condition} 
\be \lb{2}
\Om (x_A,x_B)=0 
\ee
\noindent
{\em is fulfilled}.
 
Thus, $\Om (x_A,x)=0$ is the equation of the light cone ${\cal C}_{x_{A}}$.

\section{Time transfer functions}

It follows from Property 4 that if $\Om (x_A,x_B)$ is known in ${\cal N}_{x_{A}}$, it is 
possible to determine the travel time $t_B - t_A$ of a photon connecting two 
points $x_A$ and $x_B$ as a function of $t_{A}$, $\bx_{A}$ and $\bx_{B}$ or as a 
function of $t_{B}$, $\bx_{A}$ and $\bx_{B}$. 
It must be pointed out, however, that solving the equation 
$\Om(ct_A, \bx_{A}, ct_B, \bx_{B}) = 0$ for $t_B$ 
yields two distinct solutions $t_B^{+}$ and $t_B^{-}$ since the timelike 
curve $ x^{i} = x_B^{i}$ cuts the light cone ${\cal C}_{x_{A}}$ at two points 
$x_B^{+}$ and $x_B^{-}$, $x_B^{+}$ being in the future of $x_B^{-}$. 
In the present paper, we always consider $x_A$ as the point of emission of the 
photon and $x_B$ as the point of reception, and we focus our attention on the 
determination of $t_{B}^{+} - t_{A}$ (clearly, the determination of 
$t_{B}^{-} - t_{A}$ comes within the same methodology). For the sake of brevity, we shall 
henceforth write $t_B$ instead of $t_{B}^{+}$.

In general, $t_{B} - t_{A}$ 
may be considered either as a function of the instant of emission $t_{A}$, and of $\bx_A$, 
$\bx_B$, or as a function of the instant of reception $t_{B}$ and of $\bx_A$, $\bx_B$. So 
we are led to introduce two distinct (coordinate) time transfer functions ${\cal T}_{e}$ 
and ${\cal T}_{r}$ respectively defined by 
\be \lb{2a}
t_{B} - t_A = {\cal T}_{e}(t_A, \bx_{A}, \bx_{B}) \, ,
\ee
\noindent
and
\be \lb{2aa}
t_{B} - t_A = {\cal T}_{r}(t_B, \bx_{A}, \bx_{B}) \, .
\ee 
We shall call ${\cal T}_{e}(t_A, \bx_{A}, \bx_{B})$ the emission time transfer 
function and ${\cal T}_{r}(t_B, \bx_{A}, \bx_{B})$ the reception time transfer 
function. 

There exist direct relations between the time transfer functions and the 
components of the vector tangent to a null geodesic. Indeed, it results from 
equations (\ref{2}) and (\ref{2a}) that whatever $\bx_{B}$, $x^{0}_{A}$ and $\bx_{A}$, 
one has the relation
\be \lb{2b}
\Omega\left( x^{0}_A, \bx_{A}, x^{0}_{A} + 
c{\cal T}_{e}(t_A, \bx_{A}, \bx_{B}), \bx_{B}\right) \equiv 0 \, .
\ee
\noindent
Differentiating this identity with respect to $x^{0}_{A}$, $x^{i}_{A}$ and $x^{i}_{B}$, respectively, 
it is easily seen that the relations
\be \lb{2c1}
\frac{\partial \Omega}{\partial x^{0}_{A}}(x_A,x_B) + 
\frac{\partial \Omega}{\partial x^{0}_{B}}(x_A,x_B)
\left[ 1 + \frac{\partial  {\cal T}_{e}}{\partial t_{A}}
(t_A,\bx_A,\bx_B)\right] = 0 \, ,
\ee
\be \lb{2c2}
\frac{\partial \Omega}{\partial x^{i}_{A}}(x_A,x_B) + 
c \, \frac{\partial \Omega}{\partial x^{0}_{B}}(x_A,x_B)
\, \frac{\partial  {\cal T}_{e}}{\partial x^{i}_{A}}(t_A,\bx_A,\bx_B) = 0 \, ,
\ee
\be \lb{2c3}
c \, \frac{\partial \Omega}{\partial x^{0}_{B}}(x_A,x_B) \, 
\frac{\partial  {\cal T}_{e}}{\partial x^{i}_{B}}(t_A,\bx_A,\bx_B) + 
\frac{\partial \Omega}{\partial x^{i}_{B}}(x_A,x_B) = 0 \, ,
\ee
\noindent
hold for any couple of points $(x_A, x_B)$ connected by a null geodesic. Of course, 
analogous relations may be derived from the identity
\be \lb{2ba}
\Omega\left( x^{0}_B - c{\cal T}_{r}(t_B, \bx_{A}, \bx_{B}), \bx_{A}, x^{0}_{B}, 
\bx_{B}\right) \equiv 0 \, .
\ee
Comparing these relations with equations (\ref{p1a})-(\ref{p1b}), 
we get the following theorem for the components of the vectors tangent to a light ray.

\medskip
{\bf Theorem 1}.  {\em Consider a photon emitted at point} $x_A = (ct_{A}, \bx_{A})$ 
{\em and received at point} $x_B = (ct_{B}, \bx_{B})$. {\em Denote by} $k^{\mu}$ 
{\em the vector} $dx^{\mu}/d\zeta$ {\em tangent to the photon path}, $\zeta$ {\em being any affine parameter. Then, one has relations as follow for the covariant components of the 
vector tangent} {\em at} $x_A$ {\em and} $x_B$   
\be \lb{2d1}
\left(\frac{k_{i}}{k_{0}}\right)_B = 
-c \, \frac{\partial {\cal T}_{e}}{\partial x^{i}_{B}} \, = \,
- c \, \frac{\partial  {\cal T}_{r}}{\partial x^{i}_{B}}
\left[1 - \frac{\partial  {\cal T}_{r}} {\partial t_B}\right]^{-1}\,  ,
\ee
\be \lb{2d2}
\left(\frac{k_{i}}{k_{0}}\right)_A = 
c \, \frac{\partial  {\cal T}_{e}}{\partial x^{i}_{A}}
\left[1 + \frac{\partial  {\cal T}_{e}}{\partial t_{A}}
\right]^{-1} \, = \, c \, \frac{\partial {\cal T}_{r}}{\partial x^{i}_{A}} \, ,
\ee
\be \lb{2d3}
\frac{(k_{0})_B}{(k_{0})_A} = \left[ 1 + 
\frac{\partial  {\cal T}_{e}}{\partial t_{A}}\right]^{-1} \, 
= \, 1 -
\frac{\partial  {\cal T}_{r}}{\partial t_{B}}   \, ,
\ee
\noindent
{\em where} ${\cal T}_{e}$ {\em and} ${\cal T}_{r}$ {\em are taken at} 
$(t_A, \bx_A, \bx_B)$ {\em and} $(t_B, \bx_A, \bx_B)$, {\em respectively}.  

These fundamental formulae show that all the theoretical problems related 
to the directions of light rays or to the frequency shifts may be solved 
as soon as at least one of the time transfer functions is explicitly determined. 
This property will be very useful in practice since extracting the time transfer formulae
(\ref{2a}) or (\ref{2aa}) from equation (\ref{2}), next using equations (\ref{2d1})-(\ref{2d3}) 
will be more straightforward than deriving the vectors tangent at $x_A$ and $x_B$ 
from equations (\ref{p1a})-(\ref{p1b}), next imposing constraint (\ref{2}). 

\medskip
{\em Case of a stationary space-time} -- If space-time is stationary, we can choose 
coordinates $(x^{\mu})$ such that the metric does not depend on $x^0$. Then, the world function is 
a function of $x_{B}^{0} - x_{A}^{0}$, $\bx_{A}$ and $\bx_{B}$, and the two time transfer functions 
previously introduced are identical. So  equations (\ref{2a}) and (\ref{2aa}) reduce to a single 
relation of the form 
\be \lb{4}
t_B - t_A = {\cal T} (\bx_{A}, \bx_{B}) \, .
\ee

Since ${\cal T}$ only depends on $\bx_A$ and $\bx_B$, it is immediately deduced from 
equations (\ref{2d1})-(\ref{2d3}) that the vectors $(\widetilde{k}^{\mu})_A$ and $(\widetilde{k}^{\mu})_B$ 
defined by their covariant components
\bea 
&  &(\widetilde{k}_0)_A=1, \quad (\widetilde{k}_i)_A =
c \, \frac{\partial {\cal T}}{\partial x_{A}^{i}}(\bx_A,\bx_B) \, , \lb{6} \\
&  &(\widetilde{k}_0)_B=1, \quad (\widetilde{k}_i)_B =
- \, c\, \frac{\partial {\cal T}}{\partial x_{B}^{i}}(\bx_A,\bx_B) \, , \lb{5}
\eea
are tangent to any light ray connecting $\bx_A$ and $\bx_B$, respectively. 
It must be pointed out that these tangent vectors correspond to an affine 
parameter such that $\widetilde{k}_0=1$ along the ray (note that such a parameter does 
not coincide with $\lambda$). 

\section{Determination of $\Omega (x_A,x_B)$ in the $n$th post-Minkowskian approximation}

Equations (\ref{p2b}) or (\ref{p2a}) are the basic formulae enabling to determine the 
perturbative expansion (\ref{4nn}). One might also directly use the integral form of the 
world function given by (\ref{d}), but the calculations would be very tedious (a brief outline of 
this method is given in Appendix A for the second post-Minkowskian approximation). 
 
First of all, let us briefly consider Minkowski space-time. The geodesic path connecting $x_A$ and 
$x_B$ is curve $\Gamma^{(0)}_{AB}$ having parametric 
equations $x^{\alpha} = x_{(0)}^{\alpha}(\lambda)$, with 
\be \lb{9}
x_{(0)}^{\alpha}(\lambda) = (x_{B}^{\alpha}-x_{A}^{\alpha}) \lambda +
x_{A}^{\alpha} \, , \quad 0 \leq \l \leq 1 \, . 
\ee
\noindent
According to (\ref{9}), equation (\ref{d}) gives immediately for the world function in Minkowski space-time
\be \lb{7}
\Om^{(0)}(x_A, x_B)=\frac{1}{2}\eta_{\mu \nu}(x_{B}^{\mu}-x_{A}^{\mu})
(x_{B}^{\nu}-x_{A}^{\nu}) \, .
\ee

In what follows we suppose that space-time is a perturbation of Minkowski space-time. Let us assume 
for a moment that metric (\ref{1n}) is written in the form 
\be \lb{8}
g_{\mu \nu}=\eta_{\mu \nu} + h_{\mu \nu} \, .
\ee

Henceforth we suppose that $x_B \in {\cal N}_{x_{A}}$. The geodesic curve $\Gamma_{AB}$ connecting 
$x_A$ and $x_B$ is then a 
perturbation of the straight line $\Gamma^{(0)}_{AB}$. So, the parametric equations 
of $\Gamma_{AB}$ may be written in the form
\be \lb{9a}
x^{\alpha}(\l) = (x_{B}^{\alpha}-x_{A}^{\alpha}) \lambda +
x_{A}^{\alpha} + X^{\alpha}(\l) \, , \quad 0 \leq \l \leq 1  \, ,
\ee
\noindent
where functions $X^{\alpha}$ satisfy the boundary conditions
\be \lb{9b}
X^{\alpha}(0) = 0 \, , \quad  X^{\alpha}(1) = 0 \, .
\ee
\noindent
Noting that $\dot{x}^{\mu}(\l )=x_{B}^{\mu}-x_{A}^{\mu} + \dot{X}^{\mu}(\l )$ and that
\[
\int_{0}^{1}\eta_{\mu \nu}(x_B^{\mu} - x_A^{\mu}) \dot{X}^{\nu}(\l )d\l = 0
\]
\noindent
by virtue of equations (\ref{9b}), it may be seen that equation (\ref{d}) transforms into 
\bea \lb{9c}
\fl \Om (x_A,x_B) & = & \Om^{(0)}(x_A,x_B) + \frac{1}{2} (x_B^{\mu} - x_A^{\mu})
(x_B^{\nu} - x_A^{\nu})\int_{0}^{1} h_{\mu \nu}(x(\lambda)) d\l  \nonumber \\
\fl&  & \mbox{}+ \frac{1}{2} \int_{0}^{1} \left[ \eta_{\mu\nu} \dot{X}^{\mu}(\l )
\dot{X}^{\nu}(\l ) + 
2 (x_B^{\mu} - x_A^{\mu})h_{\mu\nu}(x(\lambda)) \dot{X}^{\nu}(\l ) \right. \nonumber \\
\fl&  & \mbox{} \qquad \qquad \qquad + \left. h_{\mu \nu}(x(\lambda))
\dot{X}^{\mu}(\l ) \dot{X}^{\nu}(\l ) \right] d\l \, . 
\eea

All the integrals in (\ref{9c}) are taken over geodesic path $\Gamma_{AB}$, 
which is generally unknown. However, it is possible to obtain a perturbative 
expansion of $\Om (x_A,x_B)$ involving only line integrals taken over the straight line 
$\Gamma^{(0)}_{AB}$ when the metric is given by expansion (\ref{1n}). 
Indeed, noting that the concomitant expansion of the contravariant components $g^{\mu\nu}$ is given by 
\be \lb{1na} 
g^{\mu\nu}(x, G) = \eta^{\mu\nu} + \sum_{n=1}^{\infty}G^n \, g^{\mu\nu}_{(n)}(x) \, ,
\ee
\noindent
where the set of quantities $g^{\mu\nu}_{(n)}$ is determined by
\be \lb{1n1}
g^{\mu\nu}_{(1)} = -\eta^{\mu\rho}  \eta^{\nu\sigma} g^{(1)}_{\rho\sigma} \, ,
\ee
\be \lb{1nn}
g^{\mu\nu}_{(n)} = - \eta^{\mu\rho}  \eta^{\nu\sigma} g^{(n)}_{\rho\sigma} - 
\sum_{p=1}^{n-1}\eta^{\mu\rho} \, g^{(p)}_{\rho\sigma} \, g_{(n-p)}^{\nu\sigma} \, ,
\ee
\noindent
we can state the theorem below.

\medskip
{\bf Theorem 2}.  {\em Assuming that the metric is represented by expansion} (\ref{1n}) {\em and that} 
$x_B \in {\cal N}_{x_{A}}${\em , the world function is given by equation} (\ref{4nn}){\em , namely} 
\[
\Omega(x_A, x_B) = \Omega^{(0)}(x_A, x_B) + \sum_{n=1}^{\infty} G^n \Omega^{(n)}(x_A, x_B)  \, ,
\]
\noindent
{\em where}
\bea 
\fl& &\Omega^{(1)}(x_A, x_B) = - \frac{1}{2}\eta_{\mu\rho}\eta_{\nu\sigma}(x_{B}^{\rho}-x_{A}^{\rho})
(x_{B}^{\sigma}-x_{A}^{\sigma})\int_{0}^{1} g^{\mu\nu}_{(1)}(\x0 )d\lambda \, , \lb{4np} \\
\fl& &\Omega^{(n)}(x_A, x_B)= - \frac{1}{2}\eta_{\mu\rho}\eta_{\nu\sigma}(x_{B}^{\rho}-x_{A}^{\rho})
(x_{B}^{\sigma}-x_{A}^{\sigma})\int_{0}^{1}  g^{\mu\nu}_{(n)}(\x0) d\lambda   \nonumber \\
\fl& &  \qquad \quad - \frac{1}{2}\int_{0}^{1} \sum_{p=1}^{n-1}\, g_{(p)}^{\mu\nu}(\x0) 
\left[\frac{}{} \eta_{\mu\rho}\, (x^{\rho}_{B}-x^{\rho}_{A})\frac{1}{\lambda}\frac{\partial \Omega^{(n-p)}}{\partial x^{\nu}}(x_A,\x0 ) \right. \nonumber \\
\fl& & \qquad \qquad \qquad \left. + \sum_{q=1}^{n-p}\frac{1}{\lambda^2}\frac{\partial \Omega^{(q)}}{\partial x^{\mu}}(x_A,\x0 )
\frac{\partial \Omega^{(n-p-q)}}{\partial x^{\nu}}(x_A,\x0 ) \right]d\lambda  \nonumber \\
\fl& & \qquad \quad  -\frac{1}{2} \int_{0}^{1}\sum_{p=1}^{n-1}\frac{1}{\lambda^2}\eta^{\mu\nu} \, 
\frac{\partial \Omega^{(p)}}{\partial x^{\mu}}(x_A,\x0 ) \, 
\frac{\partial \Omega^{(n-p)}}{\partial x^{\nu}}(x_A,\x0 )d\lambda \, , \lb{4nq}
\eea
{\em whatever} $n \geq 2$. {\em All integrals are calculated along straight line} $\Gamma^{(0)}_{AB}$ 
{\em defined by equation} (\ref{9}).

Before we prove this theorem, we can enunciate a corollary as follows.

\medskip
{\bf Corollary.} {\em In terms of covariant components of the metric tensor,} $\Omega^{(1)}(x_A, x_B)$ 
{\em and} $\Omega^{(2)}(x_A, x_B)$ {\em may be written in the form}
\bea 
\fl\Omega^{(1)}(x_A, x_B) &=& \frac{1}{2}(x_{B}^{\mu}-x_{A}^{\mu})
(x_{B}^{\nu}-x_{A}^{\nu})
\int_{0}^{1} g_{\mu\nu}^{(1)}(\x0 )d\lambda \, \, , \lb{4np1} \\
\fl\Omega^{(2)}(x_A, x_B)&=& \frac{1}{2}(x_{B}^{\mu}-x_{A}^{\mu})
(x_{B}^{\nu}-x_{A}^{\nu})\int_{0}^{1} \left[ 
g_{\mu\nu}^{(2)}(\x0) \right.  \nonumber \\
\fl& & \left. \qquad \qquad \qquad \qquad  - \, \eta^{\rho\sigma}\, 
g_{\mu\rho}^{(1)}(\x0)\, g_{\nu\sigma}^{(1)}(\x0 )
\right]d\lambda  \nonumber \\
\fl& & \mbox{} +(x_{B}^{\mu}-x_{A}^{\mu})  \int_{0}^{1}\frac{1}{\lambda}
\eta^{\nu\rho}\,
\frac{\partial \Omega^{(1)}}{\partial x^{\nu}}(x_A,\x0 )\, 
g_{\mu\rho}^{(1)}(\x0 )d\lambda \nonumber \\
\fl& & \mbox{} - \frac{1}{2} \int_{0}^{1}\frac{1}{\lambda^2}\eta^{\mu\nu} \, 
\frac{\partial \Omega^{(1)}}{\partial x^{\mu}}(x_A,\x0 ) \, 
\frac{\partial \Omega^{(1)}}{\partial x^{\nu}}(x_A,\x0 )d\lambda \, . \lb{4nq2}
\eea

This corollary is an immediate consequence of equations (\ref{4np}) and (\ref{4nq}) for $n=2$ when equations (\ref{1n1})-(\ref{1nn}) are taken into account. 

\medskip
{\bf Proof of Theorem 2.} Points $x_A$ and $x_B$ being given, we may consider $\Omega(x_A, x)$ as a 
function of $x$ denoted by $\overline{\Omega}(x)$. This function satisfies the Hamilton-Jacobi 
equation (see equation (\ref{p2b}))
\be \lb{4n1}
\frac{1}{2}\, g^{\alpha \beta}(x)\frac{\partial \overline{\Omega}}{\partial x^{\alpha}}(x) 
\frac{\partial \overline{\Omega}}{\partial x^{\beta}}(x) =  \overline{\Omega} (x)\, . 
\ee

Function $\overline{\Omega} (x)$ 
may be represented by a series 
\be \lb{4n5}
\overline{\Omega} (x) = \Omega^{(0)}(x_A, x) + \sum_{n=1}^{\infty}G^n \overline{\Omega}^{(n)}(x) \, ,
\ee
\noindent
where each term $\overline{\Omega}^{(n)}$ satisfies the boundary condition
\be \lb{4n5a}
\overline{\Omega}^{(n)} (x_A) = 0 \, .
\ee
\noindent
Moreover, it follows from Property 2 that the first-order derivatives of each $\overline{\Omega}^{(n)}$ 
must satisfy the boundary condition
\be \lb{4n5b}
\lim_{ x \rightarrow x_A} \left[ \frac{\partial \overline{\Omega}^{(n)}}{\partial x^{\mu}}(x) \right] = 0 \, .
\ee
\noindent
Inserting developments (\ref{1na}) and (\ref{4n5}) into equation (\ref{4n1}), and then putting
\be \lb{4n6a}
\overline{q}_{\mu} = \frac{\partial \Omega^{(0)}}{\partial x^{\mu}}(x_A,x ) \equiv
\eta_{\mu\rho}(x^{\rho}-x_{A}^{\rho}) \, , 
\ee
\be \lb{4n6b}
\overline{q}^{\mu} = \eta^{\mu\nu}\overline{q}_{\nu} = x^{\mu} - x_{A}^{\mu} \, ,
\ee
\noindent
it is easily seen that equation (\ref{4n1}) splits up into an infinite set of partial differential equations 
as follows
\be \lb{4n7}
\overline{q}^{\mu}\frac{\partial \overline{\Omega}^{(n)}}{\partial x^{\mu}} - \overline{\Omega}^{(n)}=
\Phi^{(n)} \, ,
\ee
\noindent
where
\bea 
\fl& &\Phi^{(1)}(x) = - \frac{1}{2}\, \overline{q}_{\mu}\, \overline{q}_{\nu} \, g_{(1)}^{\mu\nu}(x) 
\, , \lb{4n8a} \\
\fl& &\Phi^{(n)}(x) = 
- \frac{1}{2}\, \overline{q}_{\mu}\, \overline{q}_{\nu}\, g_{(n)}^{\mu\nu}(x) \nonumber \\ 
\fl& & \qquad -\frac{1}{2}\, \sum_{p=1}^{n-1}\left\lbrace g_{(p)}^{\mu\nu}(x)\left[\overline{q}_{\mu} \frac{\partial \overline{\Omega}^{(n-p)}}{\partial x^{\nu}} 
+ \sum_{q=1}^{n-p}\frac{\partial \overline{\Omega}^{(q)}}{\partial x^{\mu}}
\frac{\partial \overline{\Omega}^{(n-p-q)}}{\partial x^{\nu}} \right] + \eta^{\mu\nu}
\frac{\partial \overline{\Omega}^{(p)}}{\partial x^{\mu}}
\frac{\partial \overline{\Omega}^{(n-p)}}{\partial x^{\nu}} \right\rbrace \nonumber \\
\fl& &   \lb{4n8n} 
\eea
whatever $n \geq 2$.

Assume now that $x$ moves along $\Gamma_{AB}^{(0)}$. This means that $x$ varies as a function of $\lambda$ according to equation (\ref{9}). As a consequence the total derivative of $\overline{\Omega}^{(n)}$ along $\Gamma_{AB}^{(0)}$ is given by
\be \lb{4n9}
\frac{d \overline{\Omega}^{(n)}(\x0)}{d\lambda} = (x_B^{\mu}-x_A^{\mu})
\frac{\partial \overline{\Omega}^{(n)}}{\partial x^{\mu}}
(\x0) \, ,
\ee
\noindent
which implies that equation (\ref{4n7}) may be written in the form of an ordinary differential equation
\be \lb{4n10}
\lambda \, \frac{d \overline{\Omega}^{(n)}(\x0)}{d\lambda} - \overline{\Omega}^{(n)}(\x0) = 
\Phi^{(n)}(\x0) \, .
\ee

It is easy to show that equation (\ref{4n10}) admits one and only one solution satisfying  
boundary conditions (\ref{4n5a}) and (\ref{4n5b}), namely
\be \lb{4n11}
\overline{\Omega}^{(n)}(\x0) = \lambda \, \int_{0}^{\lambda} \frac{1}{\lambda'^{2}}\, \Phi^{(n)}(x_{(0)}(\lambda'))
d \lambda' \, ,
\ee
\noindent
from which one can deduce the formulae (\ref{4np})-(\ref{4nq}). The only difficulty is to prove that $\lambda^{-2}\Phi^{(n)}(\x0)$ is a continuous function of $\lambda$ in the range 
$ 0 \leq \lambda \leq 1$. The property is obviously true for $n=1$, since it follows from equation (\ref{4n8a}) that
\be \lb{4n12} 
\fl \frac{1}{\lambda^{2}}\, \Phi^{(1)}(\x0) = -  \frac{1}{2}(x_B^{\mu} - x_A^{\mu})
(x_B^{\nu} - x_A^{\nu})\eta_{\mu\rho} \eta_{\nu\sigma}g_{(1)}^{\rho\sigma}(\x0) \, .
\ee
As a consequence, $\overline{\Omega}^{(1)}(\x0)$ is given by (\ref{4n11}) for $n=1$. Therefore, 
the formula (\ref{4np}) holds. Replacing $x_B$ by $x$ in (\ref{4np}) and then differentiating with respect 
to $x$ gives 
\bea \lb{4ns}
& &\frac{\partial \Omega^{(1)}}{\partial x^{\nu}}(x_A,x ) = - \overline{q}_{\rho}
\int_{0}^{1} \eta_{\nu\sigma}g_{(1)}^{\rho\sigma}(\overline{x}_{(0)}(\bar{\lambda})) d\bar{\lambda}  \\
& &  \qquad \qquad \qquad \quad  - \frac{1}{2}\overline{q}_{\rho}\overline{q}_{\sigma}
\int_{0}^{1} g^{\rho\sigma}_{(1) \, \, , \nu}(\overline{x}_{(0)}(\bar{\lambda}))\bar{\lambda} d\bar{\lambda} \, , \nonumber
\eea
where $\overline{x}_{(0)}(\bar{\lambda})$ is defined by 
\be \lb{4ns0}
\overline{x}_{(0)}^{\, \alpha}(\bar{\lambda}) = (x^{\alpha} - x_A^{\alpha})\bar{\lambda} + x_A^{\alpha} \, .
\ee
\noindent 
Setting $x= \x0 $ in (\ref{4ns}) yields
\bea 
\fl & &\frac{\partial \Omega^{(1)}}{\partial x^{\nu}}(x_A,\x0 ) = - \lambda (x^{\alpha}_B - x^{\alpha}_A)
\int_{0}^{1} \eta_{\alpha\rho} \eta_{\nu\sigma} g_{(1)}^{\rho\sigma}(\overline{x}_{(0)}(\bar{\lambda})) d\bar{\lambda} \nonumber \\
\fl& & \qquad \quad - \frac{1}{2} \lambda^2 (x^{\alpha}_B - x^{\alpha}_A)(x^{\beta}_B - x^{\beta}_A)
\int_{0}^{1}\eta_{\alpha\rho} \eta_{\beta\sigma} g^{\rho\sigma}_{(1)\, ,\nu}  (\overline{x}_{(0)}(\bar{\lambda}))\bar{\lambda} 
d\bar{\lambda} \, . \lb{4ns1}
\eea
Inserting this expression into equation (\ref{4n8n}) written for $n=2$, it is easily seen that 
$ \lambda^{-2}\Phi^{(2)}$ is a continuous function of $\lambda$. As a consequence, $\overline{\Omega}^{(2)}$ is 
given by (\ref{4n11}) for $n= 2$, which proves that the formula (\ref{4nq}) is valid for $n=2$. The above 
reasoning may be used to prove that $ \lambda^{-2}\Phi^{(n)}$ is continuous whatever $n$. Q. E. D.

The expression of $\Omega^{(2)}(x_A, x_B)$ given by the formula (\ref{4nq2}) leads to heavy calculations 
owing to the presence of quadratic terms with respect to the derivatives of $\Omega^{(1)}$ in one 
of the integrals. Fortunately, a drastic simplification is possible.

\medskip
{\bf Theorem 3}.  {\em The second-order term} $\Omega^{(2)}(x_A, x_B)$ {\em may be written 
in the form}
\bea \lb{4nr}
\fl& &\Omega^{(2)}(x_A, x_B)= \frac{1}{2}(x_{B}^{\mu}-x_{A}^{\mu})
(x_{B}^{\nu}-x_{A}^{\nu}) \nonumber \\ 
\fl& & \qquad \times \int_{0}^{1} \left[ \frac{}{}
g_{\mu\nu}^{(2)}(\x0) - \, \eta^{\rho\sigma}\, 
g_{\mu\rho}^{(1)}(\x0)\, g_{\nu\sigma}^{(1)}(\x0 ) \right. \nonumber \\
\fl& & \qquad \qquad \qquad \qquad \left. - \eta^{\rho\sigma}\,
\frac{\partial \Omega^{(1)}}{\partial x^{\rho}}(x_A,\x0 ) \,g_{\mu\nu , \sigma}^{(1)}(\x0 )
\right]d\lambda  \nonumber \\
\fl& & \quad  \qquad \qquad \qquad \qquad \quad + \frac{1}{2} \eta^{\mu\nu} \, 
\frac{\partial \Omega^{(1)}}{\partial x^{\mu}_{B}}(x_A,x_B ) \, 
\frac{\partial \Omega^{(1)}}{\partial x^{\nu}_{B}}(x_A,x_B ) \, , 
\eea
\noindent
{\em the integral being taken along} $\Gamma_{AB}^{(0)}$. 

\medskip
{\bf Proof of Theorem 3}. It follows from equation (\ref{4ns1}) that 
\be \lb{4nu}
\lim_{\lambda \rightarrow 0}\left[\frac{1}{\lambda}\frac{\partial \Omega^{(1)}}{\partial x^{\mu}}(x_A,\x0 )
\frac{\partial \Omega^{(1)}}{\partial x^{\nu}}(x_A,\x0 )\right] = 0 \, .
\ee 
\noindent
Taking this equation into account, an integration by parts gives
\bea 
\fl& &- \frac{1}{2} \int_{0}^{1}\frac{1}{\lambda^2}\eta^{\mu\nu} \, 
\frac{\partial \Omega^{(1)}}{\partial x^{\mu}}(x_A,\x0 ) \, 
\frac{\partial \Omega^{(1)}}{\partial x^{\nu}}(x_A,\x0 )d\lambda   \nonumber \\
\fl& &\qquad \quad  = \frac{1}{2}\frac{\partial \Omega^{(1)}}{\partial x^{\mu}_B}(x_A,x_B )
\frac{\partial \Omega^{(1)}}{\partial x^{\nu}_B}(x_A,x_B )  \nonumber \\
\fl& &\qquad \quad \; - \eta_{\mu \nu}\int_{0}^{1} \frac{1}{\lambda}\frac{\partial \Omega^{(1)}}{\partial x^{\mu}}
(x_A,\x0 ) \frac{d}{d\lambda}\left[
\frac{\partial \Omega^{(1)}}{\partial x^{\nu}}(x_A,\x0 )\right] d\lambda \, . \lb{4nv}
\eea
\noindent
In order to transform the r.h.s. of equation (\ref{4nv}), let us perform the change of variable 
$\lambda' = \lambda \bar{\lambda}$ in equation (\ref{4ns1}). We obtain
\bea \lb{4nt}
& &\frac{\partial \Omega^{(1)}}{\partial x^{\nu}}(x_A,\x0 ) = (x^{\rho}_B - x^{\rho}_{A})
\int_{0}^{\lambda} g_{\rho\nu}^{(1)}(x_{(0)}(\lambda')) d\lambda' \nonumber \\
& & \qquad \quad  + \frac{1}{2}(x^{\rho}_B - x^{\rho}_{A})(x^{\sigma}_B - x^{\sigma}_{A})
\int_{0}^{\lambda} g_{\rho\sigma ,\, \nu}^{(1)}(x_{(0)}(\lambda'))\lambda' d\lambda' \, .
\eea
Differentiating (\ref{4nt}) with respect to $\lambda$ yields 
\bea \lb{4nw}
\fl\frac{d}{d\lambda}\left[\frac{\partial \Omega^{(1)}}{\partial x^{\nu}}(x_A,\x0 )\right] &=& 
(x^{\rho}_B - x^{\rho}_{A})\, g_{\rho\nu}^{(1)}(x_{(0)}(\lambda))  \nonumber \\ 
\fl& &\mbox{} + \frac{1}{2}(x^{\rho}_B - x^{\rho}_{A})(x^{\sigma}_B - x^{\sigma}_{A})
\lambda \, g_{\rho\sigma ,\, \nu}^{(1)}(x_{(0)}(\lambda)) \, .
\eea
Substituting this expression into (\ref{4nv}), and then inserting the result of this substitution 
into (\ref{4nq}), we finally get the formula (\ref{4nr}). Q. E. D.

\section{Time transfer functions in the $n$th post-Minkowskian approximation}

It follows from equations (\ref{4nn}), (\ref{2b}) and (\ref{7}) that function ${\cal T}_{e}(t_A,\bx_a,\bx_B)$ satisfies the 
equation
\be \lb{5na}
c^2{\cal T}_{e}^{2} = R_{AB}^2 -  2\sum_{n=1}^{\infty} G^n 
\Omega^{(n)}(x_A^{0},\bx_A, x_A^{0} + c{\cal T}_{e}, \bx_B) \, ,
\ee
\noindent
where
$$
R_{AB} = \vert \bx_B - \bx_A \vert \, .
$$

Since $x_A$ is the point of emission and $x_B$ the point of reception, we are looking for the solution to equation (\ref{5na}) such that $c{\cal T}_{e}(t_A,\bx_A,\bx_B) = R_{AB} + O(G)$. Mathematically, our problem 
consists in finding a function $u(x, \varepsilon)$ governed by the equation  
\be \lb{5nb}
u^2 = x^2 + \sum_{n=1}^{\infty}\varepsilon^n h_{n}(u)
\ee
\noindent
and satisfying the condition
\be \lb{5nc}
u(x, \varepsilon)\vert_{\varepsilon =0} = x \, ,
\ee
\noindent
$\varepsilon$ being a parameter sufficiently small to warrant the convergence of the series which are handled. 
Using Maclaurin's expansion of $u$ about $\varepsilon = 0$, the solution to equation (\ref{5nb}) may be 
written in the form
\[
u(x, \varepsilon) = x + \sum_{n=1}^{\infty}\varepsilon^n u_{n}(x) \, ,
\]
\noindent
where
\[
u_{n}(x) = \frac{1}{n!}\frac{\partial^{n} u(x, \varepsilon)}{\partial \varepsilon^{n}}\bigg|_{\varepsilon =0}\; .
\]
Performing successive differentiations of (\ref{5nb}) with respect to $\varepsilon$ and then taking (\ref{5nc}) into account, a straightforward calculation yields $u_{n}(x)$ for any $n$. The first terms 
are given by
\bea 
\fl& &u_1(x) = \frac{h_{1}(x)}{2x} \, ,  \nonumber \\
\fl& &u_2(x) = \frac{1}{2x}\left[h_{2}(x) + h'_{1}(x)u_{1}(x) - u_{1}^{2}(x) \right]\, , \nonumber \\
\fl& &u_3(x) = \frac{1}{2x}\left[h_{3}(x) + h'_{2}(x)u_{1}(x) +  \frac{1}{2}h''_{1}(x) u_{1}^{2}(x) 
+ h'_{1}(x) u_{2}(x) - 2 u_{1}(x)u_{2}(x) \right] \, . \nonumber \\
\fl  \lb{5nnd}
\eea

Applying this procedure to equation (\ref{5na}), it would be possible to get ${\cal T}_{e}$ at any post-Minkowskian order. For the sake of brevity, let us define the following functions of $t_A, \bx_A, \bx_B$:
\be \lb{5ne}
\widetilde{\Omega}_{e}^{(n)}(t_A, \bx_A, \bx_B) = \Omega^{(n)}(x_A^{0}, \bx_A, x_A^{0} + R_{AB},\bx_B)
\ee
\noindent
and
\be \lb{5nf}
\widetilde{\Omega}_{e\, |\, k}^{(n)}(t_A, \bx_A, \bx_B) = \frac{\partial^{k}\Omega^{(n)}}{(\partial x_B^{0})^{k}}
(x_A^{0}, \bx_A, x_A^{0} + R_{AB},\bx_B) 
\ee
\noindent
for $k = 1, 2, 3, ...$ (be careful that $\widetilde{\Omega}_{e\, |\, k}^{(n)}$ is not a partial derivative 
of $\widetilde{\Omega}_{e}^{(n)}$). Applying equations (\ref{5nnd}), we can state the following theorem.

\medskip
{\bf Theorem 4}.  {\em In the third post-Minkowskian approximation the emission time transfer function is 
given by}
\be \lb{5c}
\fl{\cal T}_{e}(t_A,\bx_A,\bx_B) = \frac{1}{c}R_{AB} + \sum_{n=1}^{3}G^{n}{\cal T}_{e}^{(n)}(t_A,\bx_A,\bx_B) 
+ O(G^4) \, ,
\ee
\noindent
{\em where}
\bea 
\fl& &{\cal T}_{e}^{(1)} = - \frac{\widetilde{\Omega}_{e}^{(1)}}{cR_{AB}} \, , \lb{5d1} \\
\fl& &{\cal T}_{e}^{(2)} = - \frac{1}{cR_{AB}}\left[\widetilde{\Omega}_{e}^{(2)} + 
c {\cal T}_{e}^{(1)}\widetilde{\Omega}_{e \, |\, 1}^{(1)} + \frac{1}{2}c^2 ({\cal T}_{e}^{(1)})^2 \right]\, , \lb{5d2}\\
\fl& &{\cal T}_{e}^{(3)} = - \frac{1}{cR_{AB}}\left[\widetilde{\Omega}_{e}^{(3)} + 
c {\cal T}_{e}^{(1)}\widetilde{\Omega}_{e\, |\, 1}^{(2)} + \frac{1}{2}c^2 ({\cal T}_{e}^{(1)})^2
\widetilde{\Omega}_{e\, |\, 2}^{(1)} 
+ c {\cal T}_{e}^{(2)}\widetilde{\Omega}_{e\, |\, 1}^{(1)} 
+ c^2 {\cal T}_{e}^{(1)} {\cal T}_{e}^{(2)} \right] \, , \nonumber \\
\fl \lb{5d3}
\eea
{\em the functions} $\widetilde{\Omega}_{e}^{(n)}$ {\em and} $\widetilde{\Omega}_{e\, |\, k}^{(n)} $ {\em being
defined by} (\ref{5ne}) {\em and} (\ref{5nf}), {\em respectively}.

Of course, defining
\be \lb{5ng}
\widetilde{\Omega}_{r}^{(n)}(t_B, \bx_A, \bx_B) = \Omega^{(n)}(x_B^{0}- R_{AB}, \bx_A, x_B^{0},\bx_B)
\ee
\noindent
and
\be \lb{5nh}
\widetilde{\Omega}_{r\, |\, k}^{(n)}(t_B, \bx_A, \bx_B) = \frac{\partial^{k}\Omega^{(n)}}{(\partial x_A^{0})^{k}}
(x_B^{0}-R_{AB}, \bx_A, x_B^{0},\bx_B) \, ,
\ee
\noindent
a similar theorem can be stated for ${\cal T}_{r}(t_B,\bx_A,\bx_B)$.

\medskip
{\bf Theorem 4 a}.  {\em In the third post-Minkowskian approximation the reception time transfer function is 
given by}
\be \lb{5cc}
\fl{\cal T}_{r}(t_B,\bx_A,\bx_B) = \frac{1}{c}R_{AB} + \sum_{n=1}^{3}G^{n}{\cal T}_{r}^{(n)}(t_B,\bx_A,\bx_B) 
+ O(G^4) \, ,
\ee
\noindent
{\em where}
\bea 
\fl& &{\cal T}_{r}^{(1)} = - \frac{\widetilde{\Omega}_{r}^{(1)}}{cR_{AB}} \, , \lb{5f1} \\
\fl& &{\cal T}_{r}^{(2)} = - \frac{1}{cR_{AB}}\left[\widetilde{\Omega}_{r}^{(2)} - 
c {\cal T}_{r}^{(1)}\widetilde{\Omega}_{r\, |\, 1}^{(1)} + \frac{1}{2}c^2 ({\cal T}_{r}^{(1)})^2 \right]\, , \lb{5f2}\\
\fl& &{\cal T}_{r}^{(3)} = - \frac{1}{cR_{AB}}\left[\widetilde{\Omega}_{r}^{(3)} -
c {\cal T}_{r}^{(1)}\widetilde{\Omega}_{r\, |\, 1}^{(2)} + \frac{1}{2}c^2 ({\cal T}_{r}^{(1)})^2
\widetilde{\Omega}_{r\, |\, 2}^{(1)} 
- c {\cal T}_{r}^{(2)}\widetilde{\Omega}_{r\, |\, 1}^{(1)} 
+ c^2 {\cal T}_{r}^{(1)} {\cal T}_{r}^{(2)} \right] \, . \nonumber \\
\fl \lb{5f3}
\eea

It may be seen that ${\cal T}_{r}^{(1)},$ ..., ${\cal T}_{r}^{(n)}$ can be determined once 
${\cal T}_{e}^{(1)},$ ..., ${\cal T}_{e}^{(n)}$ are known and conversely. Let us examine in detail the cases where $n=1$ and $n=2$. 

Comparing expressions (\ref{5d1}) and (\ref{5f1}), we get immediately
\be \lb{5g1}
{\cal T}_{r}^{(1)}(t,\bx_A,\bx_B) = {\cal T}_{e}^{(1)}\left(t - \frac{R_{AB}}{c},\bx_A,\bx_B \right)
\ee
\noindent
and conversely
\be \lb{5h1}
{\cal T}_{e}^{(1)}(t,\bx_A,\bx_B) = {\cal T}_{r}^{(1)}\left(t + \frac{R_{AB}}{c},\bx_A,\bx_B \right) \, .
\ee

Now comparing expressions (\ref{5d2}) and (\ref{5f2}), and then noting that (\ref{5d1}) implies the identity
\bea 
\fl & &\frac{\partial {\cal T}_{e}^{(1)}}{\partial t}(t, \bx_A, \bx_B)  \nonumber  \\
\fl& &\qquad\quad = - \frac{1}{R_{AB}}
\left[ \frac{\partial \Omega^{(1)}}{\partial x_{A}^{0}}(x^{0}, \bx_A, x^{0} + R_{AB},\bx_B) + 
\frac{\partial \Omega^{(1)}}{\partial x_{B}^{0}}(x^{0}, \bx_A, x^{0} + R_{AB},\bx_B)  \right] \, , \nonumber \\
\fl \lb{5i}
\eea
\noindent
we get
\bea
\fl& &{\cal T}_{r}^{(2)}(t_B,\bx_A,\bx_B) = {\cal T}_{e}^{(2)}
\left(t_B - \frac{R_{AB}}{c},\bx_A,\bx_B\right) \nonumber \\
\fl& &\qquad \qquad - {\cal T}_{e}^{(1)}\left(t_B - \frac{R_{AB}}{c},\bx_A,\bx_B\right)\frac{\partial {\cal T}_{e}^{(1)}}{\partial t}\left(t_B - \frac{R_{AB}}{c}, \bx_A, \bx_B\right) \, . \lb{5g2}
\eea
Conversely, we have    
\bea 
\fl& &{\cal T}_{e}^{(2)}(t_A,\bx_A,\bx_B) = {\cal T}_{r}^{(2)}
\left(t_A + \frac{R_{AB}}{c},\bx_A,\bx_B\right) \nonumber \\
\fl& &\qquad \qquad + {\cal T}_{r}^{(1)}\left(t_A + \frac{R_{AB}}{c},\bx_A,\bx_B\right)\frac{\partial {\cal T}_{r}^{(1)}}{\partial t}\left(t_A + \frac{R_{AB}}{c}, \bx_A, \bx_B\right) \, . \lb{5h2}
\eea

\section{Application to a static, spherically symmetric body}

In order to apply the general results obtained in sections 4 and 5, let us determine the world function and 
the time transfer function in the second post-Minkowskian approximation. Choosing spatial isotropic coordinates, we suppose that the metric components may be written as
\begin{eqnarray}
& & g_{00}^{(1)}=-\frac{2M}{c^2r} \, , \, \quad  g_{0i}^{(1)} = 0\, , \quad g_{ij}^{(1)}= 
-2\gamma\frac{M}{c^2r}\delta_{ij} \, , \nonumber \\
& & \mbox{}  \lb{7na}\\
& & g_{00}^{(2)}= 2\beta \frac{M^2}{c^4r^2} \, ,  \quad g_{0i}^{(2)} = 0 \, , \quad g_{ij}^{(2)}=
- \frac{3}{2}\delta \frac{M^2}{c^4r^2}\delta_{ij} \, , \nonumber
\end{eqnarray}
where $\beta$ and $\gamma$ are the usual post-Newtonian parameters and $\delta$ is a post-post-Newtonian 
parameter (in general relativity, $\beta=\gamma =\delta =1$). Furthermore, we suppose that $x_A$ and 
$x_B$ are such that the connecting geodesic path is both entirely outside the body and in the weak-field region. We use 
the notations
\[ 
r= \vert \bx \vert \, , \quad   r_A = \vert \bx_A \vert \, , \quad  r_B  = \vert \bx_B \vert  \, . 
\]

It is easily seen that $G \Omega^{(1)} (x_A, x_B)$ is the term due to the mass in the multipole 
expansion of the world function given by equation (56) in \cite{linet1} (see also \cite{john1})
\be \lb{wf2m}
\fl\Omega^{(1)}(x_{A}^{0},\bx_A,x_{B}^{0},\bx_B)=
- \frac{M}{c^2} \left[ (x_{B}^{0}-x_{A}^{0})^2 +\gamma R_{AB}^{2}\right] F(\bx_A,\bx_B) \, ,
\ee
\noindent
where 
\be \lb{deff0}
F(\bx_A,\bx_B) = \int_{0}^{1}\frac{d\lambda}{\vert \bx_{(0)}(\lambda )\vert} = \frac{1}{R_{AB}}
\ln \left( \frac{r_A+r_B+R_{AB}}{r_A+r_B-R_{AB}}\right) \, .
\ee

For the integrals involving $g_{\mu\nu}^{(2)}$ and terms quadratic in $g_{\mu\nu}^{(1)}$ in the 
expression of $\Omega^{(2)}(x_A,x_B)$, we find 
\bea
\fl& &\frac{1}{2}(x_{B}^{\mu}-x_{A}^{\mu})(x_{B}^{\nu}-x_{A}^{\nu}) \int_{0}^{1} \left[ g_{\mu\nu}^{(2)}
(\x0) - \, \eta^{\rho\sigma}\, g_{\mu\rho}^{(1)}(\x0)\, g_{\nu\sigma}^{(1)}(\x0 ) \right]d\lambda  \nonumber \\
\fl& & \qquad \qquad = \frac{M^2}{c^4} \left[(\beta -2)(x_B^0-x_A^0)^2 + \left(2\gamma^2 -\frac{3\delta}{4}\right)R_{AB}^2
\right] E(\bx_A, \bx_B) \, , \lb{7nc}
\eea
where $E(\bx_A, \bx_B)$ is defined as
\[ 
E(\bx_A, \bx_B)  =  \int_{0}^{1}\frac{d\lambda}{\vert \bx_{(0)}(\lambda)\vert^2} \, .
\]
\noindent
An elementary quadrature yields  
\bea 
\fl E(\bx_A, \bx_B)&=& \frac{1}{\sqrt{r_A^2 \, r_B^2-(\bx_A\cdot \bx_B)^2}}\left\lbrace \arctan \left[\frac{(\bx_B - \bx_A)\cdot \bx_B}{\sqrt{r_A^2 \, r_B^2-(\bx_A\cdot \bx_B)^2}} \right] \right.\nonumber \\
\fl& & \mbox{} \left. \qquad \quad- \arctan \left[\frac{(\bx_B - \bx_A)\cdot \bx_A}{\sqrt{r_A^2 \, r_B^2-(\bx_A\cdot \bx_B)^2}}   \right]\right\rbrace  \nonumber \\ 
\fl&=& \frac{\arccos(\bn_A\cdot \bn_B)}{r_Ar_B\sqrt{1-(\bn_A\cdot \bn_B)^2}} \, , \lb{7nd}
\eea
\noindent
$\bn_A$ and $\bn_B$ being defined as
\[
\bn_A=\bx_A/r_A \, ,  \qquad  \bn_{B}=\bx_B/r_B \, .
\]

For the integral involving the gradient of $\Omega^{(1)}$ in the formula (\ref{4nr}), we obtain 
\bea
\fl& & - \,\frac{1}{2}(x_{B}^{\mu}-x_{A}^{\mu})(x_{B}^{\nu}-x_{A}^{\nu}) \int_{0}^{1}  
\eta^{\rho\sigma}\, \frac{\partial \Omega^{(1)}}{\partial x^{\rho}}(x_A,\x0 ) \, 
g_{\mu\nu , \sigma}^{(1)}(\x0 )d\lambda  \nonumber \\
\fl& & \qquad \quad = \frac{M^2}{c^4}\left[(x_B^0-x_A^0)^2 + \gamma R_{AB}^2 \right]\left[ 
\left( 1 + \frac{r_A}{r_B} \right) \frac{1}{R_{AB}^2} \frac{(x_B^0-x_A^0)^2 + \gamma R_{AB}^2 }{r_A r_B + (\bx_A \cdot \bx_B)} \right. \nonumber \\
\fl& & \qquad \quad \qquad \qquad \qquad \left. - 2 \gamma E(\bx_A, \bx_B) 
- \frac{(x_B^0-x_A^0)^2 - \gamma R_{AB}^2 }{R_{AB}^2}\frac{F(\bx_A, \bx_B)}{r_B}  \right] \, . \nonumber \\
\fl   \lb{7nf}
\eea

Substituting (\ref{7nc}) and (\ref{7nf}) into (\ref{4nr}), and then carrying out 
the calculation of the square of the gradient of $\Omega^{(1)}(x_A, x_B)$ at $x_B$, we obtain an expression for the 
world function as follows 
\begin{eqnarray}
\fl& &\Omega(x_A,x_B)=\frac{1}{2}(x_B^0-x_A^0)^2-\frac{1}{2}R_{AB}^2-\frac{GM}{c^2}\left[
(x_B^0-x_A^0)^2+\gamma R_{AB}^2\right]F(\bx_A, \bx_B)\nonumber\\
\fl& &\qquad \quad +\frac{G^2M^2}{c^4}\left\lbrace \frac{(x_B^0-x_A^0)^4}{R_{AB}^2}\left[\frac{1}{r_A r_B + (\bx_A \cdot \bx_B)}-\frac{1}{2}F^2(\bx_A, \bx_B)\right]\right. \nonumber \\
\fl& &\qquad \quad  +(x_B^0-x_A^0)^2\left[\frac{2\gamma}{r_A r_B + (\bx_A \cdot \bx_B)}-(2-\beta+2\gamma)E(\bx_A,\bx_B) \right.  \nonumber \\
\fl& &\qquad \quad \left. \qquad  \qquad \qquad \qquad \qquad \qquad  + \frac{}{}(2+\gamma)F^2(\bx_A,\bx_B)
\right]\nonumber \\
\fl& &\qquad \quad \left. +R_{AB}^2\left[\frac{\gamma^2}{r_A r_B + (\bx_A \cdot \bx_B)}-\frac{3\delta}{4} E(\bx_A, \bx_B)-\frac{\gamma^2}{2}F^2(\bx_A,\bx_B)
\right]\right\rbrace + O(G^3) \, . \nonumber \\
\fl   \lb{7ng}
\end{eqnarray}

Now, using the formulae (\ref{5d1}) and (\ref{5d2}), a straightforward calculation gives for the time transfer function 
\begin{eqnarray}
\fl& & \mathcal{T}(\bx_A,\bx_B)  = \frac{R_{AB}}{c} +\frac{(\gamma+1)GM}{c^{3}}
\ln\left(\frac{r_A+r_B+R_{AB}}{r_A+r_B-R_{AB}}\right)\nonumber \\
\fl& & +\frac{G^{2}M^{2}R_{AB}}{c^{5}}\left[
\frac{(8-4\beta+8\gamma+3\delta)\arccos (\bn_A\cdot \bn_B)}
{4\sqrt{r_A^2 \, r_B^2-(\bx_A\cdot \bx_B)^2}}
-\frac{(1+\gamma)^2}{r_A \, r_B + (\bx_A \cdot \bx_B)}\right] + O(G^3) \, . \nonumber \\
\fl & & \lb{7n}
\end{eqnarray}

The first-order term is the Shapiro time delay \cite{shapiro}. The second-order terms in (\ref{7n}) 
generalize the result found by Brumberg in the case of general relativity \cite{brumberg}.

Denoting by $\psi_{A B}$ the angle formed by vectors $\bn_A$ and $\bn_B$ and noticing that $E(\bx_A, \bx_B)$ reads 
\be \lb{7nh}
\fl\frac{\arccos (\bn_A\cdot \bn_B)}{\sqrt{r_A^2 \, r_B^2-(\bx_A\cdot \bx_B)^2}} = 
\frac{\psi_{ A B}}{r_A r_B \sin \psi_{A B}} \, ,   \quad  0\leq \psi_{A B} < \pi \, ,
\ee
\noindent
we find an expression for $\mathcal{T}(\bx_A,\bx_B)$ as follows
\bea 
\fl& & \mathcal{T}(\bx_A,\bx_B)  = \frac{R_{AB}}{c} \left\{ 1+\frac{(\gamma+1)GM}{c^{2}R_{AB}}
\ln\left(\frac{r_A+r_B+R_{AB}}{r_A+r_B-R_{AB}}\right) \right. \nonumber \\
\fl& &\qquad \left. +\frac{G^{2}M^{2}}{c^{4}r_A r_B}\left[
(2-\beta+2\gamma+\frac{3}{4}\delta)\frac{\psi_{A B}}{\sin \psi_{A B}}
-\frac{(1+\gamma)^2}{1 + \cos \psi_{A B}}\right] \right\} + O(G^3) \, , \lb{7ni}
\eea
which may be very helpful for numerical estimates.

\section{Conclusion}

The central result of this paper is given by Theorem 2. Assuming that $x_B$ is in the normal convex 
neighbourhood of $x_A$, we have shown that $\Omega(x_A, x_B)$ can be 
obtained in the $n$th post-Minkowskian approximation by a recursive procedure which spares the trouble 
of solving the geodesic equations. Any $n$th-order perturbation $\Omega^{(n)}(x_A, x_B)$ is an integral 
taken along the zeroth-order straight line joining $x_A$ and $x_B$. Moreover, in Theorem 3 we have found 
a remarkable simplification of the integral giving $\Omega^{(2)}(x_A,x_B)$. In Theorems 4 and 4a we have 
outlined a recursive procedure enabling to obtain the perturbations of the time transfer functions when the 
perturbations of $\Omega (x_A,x_B)$ are known, assuming that the null cone at $x_A$ is included in the 
normal convex neighbourhood of $x_A$. Since the time transfer functions are sufficient to determine 
the direction of a light ray in the 3-space relative to a given observer (see Theorem 1), the 
systematic methods explored here seem very promising to tackle the relativistic problems raised 
by highly accurate astrometry in space.

Using the simplification found in the second post-Minkowskian approximation, we have shown that 
the calculation of $\Omega^{(2)}(x_A,x_B)$ and ${\cal T}^{(2)}(\bx_A,\bx_B)$ reduces to elementary 
integrations in the case of a static, spherically symmetric space-time. The simplification within 
the second order works out so well that the question whether it works also for the higher-order terms 
naturally arises. We hope to return to this problem later. For the time being, we are applying 
the results obtained in the static, spherically symmetric case to the description of the 
light deflection between two points located at a finite distance.

\appendix
\section{Another determination of $\Omega^{(2)}(x_A, x_B)$}

It is possible to derive $\Omega^{(2)}(x_A, x_B)$ directly from the Euler-Lagrange equations 
of geodesics, namely
\be \lb{2n}
\frac{d}{d\lambda}\left( (\eta_{\rho \sigma} + h_{\rho \sigma})\frac{dx^{\sigma}}{d\lambda}
\right) =\frac{1}{2}h_{\mu \nu ,\rho}
\frac{dx^{\mu}}{d\lambda}\frac{dx^{\nu}}{d\lambda} \, .
\ee

It is clear that the functions $X^{\mu}$ defined by equations (\ref{9a}) admit the following 
expansions
\be \lb{2na}
X^{\mu}(\lambda ,G)= \sum_{n=1}^{\infty}G^{n}X_{(n)}^{\mu}(\lambda) \, ,
\ee
\noindent
where the functions $X_{(n)}^{\mu}(\lambda)$ satisfy the boundary conditions
\be \lb{9b}
X^{\mu}_{(n)}(0) = 0 \, , \quad  X^{\mu}_{(n)}(1) = 0 \, \quad \forall n \geq 1.
\ee

Substituting for $g_{\mu\nu}$ and $X^{\mu}$ from equations (\ref{1n}) and (\ref{2na}) 
respectively into equation (\ref{9c}) gives
\bea 
\fl\Omega (x_A, x_B) &=&  \Omega^{(0)}(x_A, x_B) + 
\frac{1}{2}G(x_{B}^{\mu}-x_{A}^{\mu})(x_{B}^{\nu}-x_{A}^{\nu})
\int_{0}^{1} g_{\mu\nu}^{(1)}(x(\lambda))d\lambda \nonumber \\
\fl& & + \frac{1}{2}G^2 \int_{0}^{1}
\left[ (x_{B}^{\mu}-x_{A}^{\mu})(x_{B}^{\nu}-x_{A}^{\nu})g_{\mu\nu}^{(2)}(\x0 ) \right. \nonumber \\
\fl& &\left. + 2(x_{B}^{\mu} - x_{A}^{\mu})\dot{X}^{\nu}_{(1)}(\l )
g_{\mu\nu}^{(1)}(\x0 ) + \eta_{\mu\nu}\dot{X}^{\mu}_{(1)}(\l )\dot{X}^{\nu}_{(1)}(\l )
\right] d\lambda + O(G^3) \, . \nonumber \\
\fl \lb{4n}
\eea
Noting that
\be \lb{4na}
\fl g_{\mu\nu}^{(1)}(x(\l ) ) = 
g_{\mu\nu}^{(1)}(\x0 ) + GX^{\rho}_{(1)}g_{\mu\nu , \rho}^{(1)}(\x0 ) 
+ O(G^2) 
\ee
\noindent 
we find that $\Omega^{(1)}$ and $\Omega^{(2)}$ are given by
\bea 
\fl& &\Omega^{(1)}(x_A, x_B) = \frac{1}{2}(x_{B}^{\mu}-x_{A}^{\mu})(x_{B}^{\nu}-x_{A}^{\nu})
\int_{0}^{1} g_{\mu\nu}^{(1)}(\x0 )d\lambda \, , \lb{4nc} \\ 
\fl& &\Omega^{(2)}(x_A, x_B)= \frac{1}{2}\int_{0}^{1} \left\lbrace (x_{B}^{\mu}-x_{A}^{\mu})(x_{B}^{\nu}-x_{A}^{\nu}) \left[ g_{\mu\nu}^{(2)}(\x0 )  
+X^{\rho}_{(1)}(\l )g_{\mu\nu ,\rho}^{(1)}(\x0 )\right] \right. \nonumber \\
\fl&  &  \qquad \qquad \qquad \left. +2(x_{B}^{\mu} - x_{A}^{\mu})\dot{X}^{\nu}_{(1)}(\l )
g_{\mu\nu}^{(1)}(\x0 ) 
+ \eta_{\mu\nu}\dot{X}^{\mu}_{(1)}(\l )\dot{X}^{\nu}_{(1)}(\l )\right\rbrace d\lambda \, , \nonumber \\
\fl \lb{4nd}
\eea
where the integrals are now calculated along unperturbed geodesic $\Gamma^{(0)}_{AB}$.

Equation (\ref{4nc}) coincides with the formula (\ref{4np1}). Let us now transform the r.h.s. of equation (\ref{4nd}) 
in order to get rid of all the terms involving $X^{\mu}_{(1)}$ or $\dot{X}_{(1)}^{\mu}$. Using 
equations (\ref{2n}) and retaining only the terms of first order in $G$, we obtain
\[ 
\fl\frac{1}{2}(x_{B}^{\mu}-x_{A}^{\mu})(x_{B}^{\nu}-x_{A}^{\nu})g_{\mu\nu ,\rho}^{(1)}(\x0 ) = 
\frac{d}{d\lambda}\left[\eta_{\rho\sigma}\dot{X}_{(1)}^{\sigma}(\l ) + (x_{B}^{\sigma} - x_{A}^{\sigma})
g_{\rho\sigma}^{(1)}(\x0 ) \right] \, . \nonumber \\
\]
\noindent
Substituting the r.h.s. of these equations into equation (\ref{4nd}), and then integrating by parts, we find   
\bea  
\fl& &\Omega^{(2)}(x_A, x_B) = \frac{1}{2}\int_{0}^{1}\left[(x_{B}^{\mu}-x_{A}^{\mu})(x_{B}^{\nu}-x_{A}^{\nu}) 
g_{\mu\nu}^{(2)}(\x0 ) - \eta_{\mu\nu}\dot{X}^{\mu}_{(1)}(\l )\dot{X}^{\nu}_{(1)}(\l )\right] d\lambda 
\, . \nonumber \\
\fl& &  \lb{4nf} 
\eea
\noindent
In order to transform the integral
\be \lb{4ng}
\Xi_{(2)} = \frac{1}{2}\int_{0}^{1} \eta_{\mu\nu}\dot{X}^{\mu}_{(1)}(\l )\dot{X}^{\nu}_{(1)}(\l )d\lambda \, ,
\ee
\noindent 
let us insert expansions (\ref{2na}) and equation (\ref{4nn}) into equations (\ref{q3a}). Retaining only the terms of first order in $G$, we find that $X_{(1)}^{\mu}$ 
satisfies the differential equation 
\begin{equation}
\label{diffEq}
\fl\dot{X}^{\mu}_{(1)}-\frac{1}{\lambda}X_{(1)}^{\mu}=
\frac{1}{\lambda}\eta^{\mu\rho}
\frac{\partial \Omega^{(1)}}{\partial x^{\rho}}(x_A,x_{(0)}(\l ))
-\eta^{\mu\rho}(x_{B}^{\sigma}-x_{A}^{\sigma})g_{\rho\sigma}^{(1)}(\x0 ) \, .
\end{equation}
\noindent
The only solution to this equation satisfying the boundary condition $X_{(1)}^{\mu}(1) = 0$ is given by
\be \lb{4nl}
X_{(1)}^{\mu}(\lambda) = \lambda Y_{(1)}^{\mu} (\lambda) \, ,
\ee
\noindent
where 
\bea
\fl & &Y_{(1)}^{\mu}(\lambda)= \eta^{\mu\rho}\int_{\lambda}^{1}
\left[  \frac{1}{\lambda '}(x_{B}^{\sigma}-x_{A}^{\sigma})g_{\rho\sigma}^{(1)}(x_{(0)}(\lambda '))
- \frac{1}{\lambda '^{2}}\frac{\partial \Omega^{(1)}}{\partial x^{\rho}}
(x_A,x_{(0)}(\lambda ')) \right]d\lambda ' \, . \nonumber \\
\fl & & \lb{4nm}
\eea
\noindent
Using Property 2, it is easily checked that this solution also satisfies the boundary condition 
$X_{(1)}^{\mu}(0) = 0$.

Now substituting for $\dot{X}_{(1)}^{\mu}(\lambda)$ from 
$\dot{X}_{(1)}^{\mu}(\lambda )=\lambda \dot{Y}^{\mu}_{(1)}(\lambda )+Y^{\mu}_{(1)}(\lambda )$ 
into equation (\ref{4ng}) and then integrating by parts, we get 
\be \lb{4nnm}
\Xi_{(2)} = \frac{1}{2}\int_{0}^{1}\lambda^2 
\eta_{\mu\nu} \dot{Y}_{(1)}^{\mu}(\lambda )\dot{Y}_{(1)}^{\nu}(\lambda )d\lambda \, ,
\ee
\noindent
since $Y_{(1)}^{\mu}(1) = 0$. Differentiating the r.h.s. of equations (\ref{4nm}) with respect to $\lambda$ and then 
substituting into equation (\ref{4nnm}), we finally obtain
\begin{eqnarray}
\Xi_{(2)}&=& \frac{1}{2}(x_{B}^{\mu}-x_{A}^{\mu})
(x_{B}^{\nu}-x_{A}^{\nu})\int_{0}^{1}
\eta^{\rho\sigma}g_{\mu\rho}^{(1)}(\x0 )g_{\nu\sigma}^{(1)}(\x0 )d\lambda \nonumber \\
& & \mbox{} -(x_{B}^{\mu}-x_{A}^{\mu})  \int_{0}^{1}\frac{1}{\lambda}\eta^{\nu\rho}
\frac{\partial \Omega^{(1)}}{\partial x^{\nu}}(x_A,\x0 )g_{\mu\rho}^{(1)}(\x0 )d\lambda \nonumber\\
& & \mbox{}+ \frac{1}{2} \int_{0}^{1}\frac{1}{\lambda^2}\eta^{\mu\nu}
\frac{\partial \Omega^{(1)}}{\partial x^{\mu}}(x_A,\x0 ) \, 
\frac{\partial \Omega^{(1)}}{\partial x^{\nu}}(x_A,\x0 )d\lambda \, . \nonumber 
\end{eqnarray}
\noindent
Substituting for $\Xi_{(2)}$ from this last equation into equation (\ref{4nf}), we recover the 
expression of $\Omega^{(2)}(x_A, x_B)$ given by the formula (\ref{4nq2}). 

As a final remark, we shall note that equations (\ref{4nl}) and (\ref{4nm}) 
yield the first-order perturbative term of the geodesic path joining two given 
points $x_A$ and $x_B$ as integrals taken over the straight line with ends at 
$x_A$ and $x_B$.

\end{document}